\documentclass{article}
\usepackage{amsmath}
\usepackage{amsfonts}
\usepackage{amssymb}

\begin{document}

\title{Neutrinospheres, resonant neutrino oscillations, and pulsar kicks}
\author{M. Barkovich$^\dagger$, J. C. D'Olivo$^\dagger$, and R. Montemayor$^{*}$}
\date{}
\maketitle
\begin{flushleft}
$\dagger$ Departamento de F\'{\i}sica de Altas Energ\'{\i}as,
Instituto de Ciencias Nucleares, Universidad Nacional Aut\'{o}noma
de M\'{e}xico, Apartado Postal 70-543, 04510 M\'{e}xico, Distrito
Federal, M\'{e}xico. \vskip 0.3cm $*$ Instituto Balseiro and
Centro At\'omico Bariloche, Universidad Nacional de Cuyo and CNEA,
8400 S. C. de Bariloche, R\'{\i}o Negro, Argentina.
\end{flushleft}

\begin{abstract}

Pulsars are rapidly rotating neutron stars and are the outcome of
the collapse of the core of a massive star with a mass of the
order of or larger than eight solar masses. This process releases
a huge gravitational energy of about $10^{53}\,\rm{erg}$, mainly
in the form of neutrinos. During the collapse the density
increases, and so does the magnetic field due to the trapping of
the flux lines of the progenitor star by the high conductivity
plasma. When the density reaches a value of around
$10^{12}\,\rm{g}\,\rm{cm}^{-3}$ neutrinos become trapped within
the protoneutron star and a neutrinosphere, characterized inside
by a diffusive transport of neutrinos and outside by a free
streaming of neutrinos, is formed and lasts for a few seconds.
Here we focus on the structure of the neutrinosphere, the
resonant flavor conversion that can happen in its interior, and
the neutrino flux anisotropies induced by this phenomena in the
presence of a strong magnetic field. We present a detailed
discussion in the context of the spherical Eddington model, which
provides a simple but reasonable description of a static neutrino
atmosphere, locally homogenous and isotropic. Energy and momentum
are transported by neutrinos and antineutrinos flowing through an
ideal gas of nonrelativistic, nondegenerate nucleons and
relativistic, degenerate electrons and positrons. We examine the
details of the asymmetric neutrino emission driven by
active-sterile neutrino oscillations in the magnetized
protoneutron star, and the possibility for this mechanism to
explain the intrinsic large velocities of pulsars respect to
nearby stars and associated supernova remnants.

\end{abstract}

\newpage
\section{Introduction}

The finding of pulsars in the sixties was one of the most
astonishing discoveries in astrophysics\cite{discover}. Soon
after their observation they were identified with neutron
stars\cite{ident}, whose existence had been theoretically
predicted decades before. Neutron stars are the densest, most
rapidly rotating, and most strongly magnetized objects in the
Galaxy, features well understood by the standard physics of
gravitational and elementary particle physics\cite{ST}. They are
unique laboratories, which give us access to extreme conditions
that are virtually impossible to obtain on Earth. Neutron stars
are of particular interest for neutrino physics because of the
key role that these particles play in the very first stages of
their formation\cite{raffelt}.

There is a remarkable characteristic of pulsars not
satisfactorily understood: their large drift velocities with
respect to nearby stars. The average value
of pulsar velocities is in the range of $200-500\,\mathrm{km}%
\,\mathrm{s}^{-1}$, an order of magnitude larger than the mean
velocity of ordinary stars in our Galaxy
($\sim30\,\mathrm{km}\,\mathrm{s}^{-1}$)\cite{pulsar}. In
addition, the distribution of these velocities is not Gaussian.
It seems to have a bimodal structure\cite{bimodal}, with a
significant fraction ($\sim15\%$) of the pulsar population having
velocities higher than $1000\,\mathrm{km}\,\mathrm{s}^{-1}$ and up
to a maximum of $1600\,\mathrm{km}\,\mathrm{s}^{-1}$.

Neutron stars are formed in type-II supernova explosions and,
given the enormous energy liberated during these processes, it is
quite natural to look for an explanation in terms of an impulse or
\textit{kick} received during their birth. This hypothesis is
supported by recent analysis of individual pulsar motions and of
associations between supernovae remnants and pulsars. It is also
consistent with observed characteristics of binary systems with
one or both constituents being a neutron star\cite{natal}.
Despite this evidence, the physical origin of the kick is one of
the most important unsolved puzzles in supernovae research. If the
distribution is bimodal, then more than one mechanism could be
responsible for the natal kicks. Two classes of mechanism have
been considered\cite{DL}. One of them invokes an asymmetric mass
ejection during the supernova explosion as the result of
hydrodynamic perturbations caused by global density
inhomogeneities prior to the core collapse\cite{global} or
convective instabilities in the neutrino-heated layer behind the
shock\cite{convect}. Recoil speeds as high as
$\sim500\,\mathrm{km}\,\mathrm{s}^{-1}$ have been obtained with
two-dimensional simulations. However, a recent three-dimensional
calculation shows that even the most extreme asymmetric collapse
does not produce final neutron star velocities above
$200\,\mathrm{km}\,\mathrm{s}^{-1}$\cite{fryer}, rendering the
situation unclear.

Only a small fraction of the gravitational binding energy
($E\simeq10^{53}\,\mathrm{erg}$) liberated during a supernova
explosion is required to account for the huge electromagnetic
luminosity observed and the ejection of the envelope. More than
$99\%$ of the energy is released in the form of neutrinos, which
are copiously produced when the iron core turns into a neutron
core. A small asymmetry ($\sim1\%$) in the momentum taken away by
neutrinos should be enough to give the nascent pulsar an
acceleration consistent with the measured velocities. The
asymmetric neutrino emission induced by strong magnetic fields is
the basis of the second class of kick mechanisms, and effects such
as parity violation\cite{AL}, asymmetric field
distributions\cite{LQ}, and dark spots\cite{DT} have been
considered in this context.

The density in the interior of a protoneutron star is so high, of
the order of the nuclear density, that even neutrinos remain
trapped, being emitted from an approximately well defined surface,
the neutrinosphere. In a magnetized stellar plasma there are
anisotropic contributions to the neutrino refraction
index\cite{DNEC,DN} and, as a consequence, the resonant flavor
transformations of neutrinos diffusing in different directions
respect to the magnetic field occur at different depths within the
protostar. This phenomenon can induce a momentum flux asymmetry
and is the basis of the neutrino-driven kick mechanism proposed by
Kusenko and Segr\`{e} (KS)\cite{KS,otros,BDMZ}. With standard
(active) neutrinos the mechanism works when the resonance region
lies between the $\nu_{e}$ and $\nu_{\mu}$ (or $\nu_{\tau}$)
neutrinospheres, but requires an exceedingly high square mass
difference, in conflict with the existing limits. A variant
avoiding such a limitation can be implemented in terms of
active-sterile neutrino oscillation\cite{KSS,KS1,n3}.

The measurement of the Z-boson width at LEP, limiting to three the
number of light neutrinos having weak interactions, does not
preclude the existence of additional neutral fermions that do not
fill the standard model gauge interactions. They have been invoked
to explain several questions in astronomy and cosmology\cite{a&c}
and may becomes a necessity if the claimed LSND evidence for
oscillations were confirmed. The range of the oscillation
parameters (mass and mixing angle) required by the pulsar kick
mechanism overlaps with the allowed region for sterile neutrinos
being candidates to cosmological dark matter. Alternative kick
mechanisms driven by neutrino oscillations have been proposed
invoking the possible existence of transition magnetic
moments\cite{momen}, non orthonormality of the flavor
neutrinos\cite{valle}, violations of the equivalence
principle\cite{vep}, or off-resonant emission of sterile
neutrinos\cite{KS1}.

In this article we review the explanation of the pulsar kick in
terms of the resonant neutrino oscillations. In Section II, we
outline the creation of a protoneutron star as a result of a
supernova event, and the formation in its interior of a gas of
trapped neutrinos. The momentum flux carried by neutrinos is
analyzed in Section III. Section IV is devoted to the discussion
of the neutrino oscillations in matter incorporating the effect of
strong magnetic fields. The implication of this phenomenon as
source of the asymmetry in the momentum flux transported by
neutrinos emitted during the protoneutron star cooling is
examined in Section V. Numerical results are obtained in Section
VI within the context of the Eddington model, where the kick
effect is explicitly evaluated. In the final section we summarize
the outstanding features of the mechanism.

\section{Stellar collapse, protoneutron stars, and neutrinospheres}

Neutron stars are extremely compact objects born in type II
supernova explosions\cite{bethe}, as the aftermath of the
gravitational collapse of a massive star ($M\gtrsim8M_{\odot})$
when it exhausts the sources of thermonuclear energy and its iron
core reaches the Chandrasekhar limit, $M_{Ch}=1.44\,M_{\odot}$.
The process is triggered by the photodissociation of iron nuclei,
$\gamma\,+\,^{56}\text{Fe} \rightarrow13\,^{4}\text{He}\,+\,4n
\,-\,124,4$\,MeV, which consumes energy reducing the pressure of
the degenerated electron gas. When the collapse proceeds, pressure
support decreases because of the electron capture by nuclei.
Electron neutrinos are copiously produced through the reaction
$p\,+\,e \rightarrow n\,+\,\nu_{e}$ and initially they freely
escape from the core, but as density increases the medium becomes
less transparent for them. The inner core ($M
\approx0.5-0.8\,M_{\odot}$) collapses in a homologous way at a
subsonic velocity proportional to the radius
($v/r=\,400-700\,\text{s}^{-1})$, with the density and temperature
profiles remaining similar and only the scale changing with
time\cite{goldreich}. The external shells collapse at a supersonic
velocity of the order of free fall velocity
($v\varpropto1/\sqrt{r}$).

During the first part of the collapse the main source of opacity
for neutrinos is the coherent scattering by heavy nuclei,
mediated by the neutral current weak interaction. The cross
section for this reaction is proportional to $A^{2}$, where $A$
is the number of nucleons in the nuclei, and is very effective in
trapping neutrinos\cite{freedman}. If $\lambda_{\nu}$ is the mean
free path of neutrinos, then the time for them to diffuse out of
the core, estimated from a random walk through a distance of the
order of the core radius $R$, is
$t_{diff}\sim{3R^{2}}/{\lambda_{\nu}}$. On the other hand, the
core dynamical time scale is of the order of the free fall time
$t_{c}\sim(G\rho)^{-1/2}$. At a certain density, of about $10^{12}%
\,\text{g\,cm}^{3}$ for 10 MeV neutrinos, both times become
comparable and neutrinos are trapped\cite{maz1}.
$\beta$-equilibrium between $e$ and $\nu_{e}$ is set up and
electron neutrinos also become degenerated. In this way, the
collapsing core is the only place, besides the early universe,
where neutrinos are in thermal equilibrium. During the collapse
the lepton fraction $Y_{L}= Y_{e}\,+\,Y_{\nu_{e}}$ is kept almost
constant near the value of the electron fraction at the beginning
of the trapping ($Y_{e} \approx0.35$), with only $\sim1/4$
corresponding to neutrinos\cite{raffelt}.

The degenerate gas of trapped neutrinos diffuses towards the
exterior, reaching regions of decreasing density where the
collisions are less frequent, until they finally decouple from the
star. Although this process is continuous, it is convenient to
introduce an effective emission surface called the neutrinosphere,
characterized as the surface where the optical depth becomes
approximately one. More accurately, the radius of the
neutrinosphere $R_{\nu_{\ell}}$ is defined by the condition
\begin{equation}
\int_{R_{\nu_{\ell}}}^{\infty}\frac{dr}{\lambda_{\nu_{\ell}}}\sim\frac{2}
{3}\,.\label{meanfp}
\end{equation}
The medium is considered opaque for neutrinos inside the neutrinosphere and
transparent outside. We can also define the energy sphere, where the energy
exchanging reactions freeze out while energy conserving collisions may still
be important.

The collapse stops suddenly when, after a fraction of seconds,
densities of around $3\times10^{14}\,\text{g\,cm}^{-3}$, exceeding
nuclear density, are reached\cite{CJ}. Then, a shock wave is
built up near the interface of the halted inner core and the
supersonically free falling outer part. The shock wave propagates
outwards and eventually ejects the stellar mantle of the star. At
its passage through the neutrinosphere nuclei are dissociated
decreasing the neutrino opacity and causing a recession of the
neutrinosphere. The $\nu_{e}$ produced by the reaction $p\,+\,e
\rightarrow n+ \nu_{e}$ can escape freely building up a short
($<$\,10\,msec) $\nu_{e}$ burst, called the ``neutronization
burst" or ``deleptonization burst". The star remnant
gravitationally decouples from the expanding ejecta and a
protoneutron star has been formed\cite{BL}.

The neutrinos continue trapped due to the interaction with the
residual particles of the medium, mainly nucleons, electrons, and
neutrinos. For $\nu_{e}$ the dominant reactions are the $\beta$
and $\beta$-inverse, which mantain them in local thermodynamical
equilibrium\cite{BM}. In this case, the distinction between
neutrinosphere and energy sphere is not crucial. The electron
neutrinos also interact with nucleons through the neutral current,
but the cross sections for these reactions are smaller. Although
the energy dependence of the process involving $\nu_{e}$ and
$\bar{\nu}_{e}$ is the same, in a protoneutron star the number of
neutrons is larger than the number of protons, and therefore the
$\nu_{e}$ suffer more collisions than the $\bar{\nu}_{e}$. As a
consequence, the neutrinosphere of the $\nu_{e}$ is a bit further
out than the one of the $\bar{\nu}_{e}$, and the mean energy per
emitted particle is smaller. After the achievement of $\beta$
equilibrium, the total entropy is conserved and the collapse
becomes adiabatic.

In the case of the nonelectron (active) neutrinos and
antineutrinos, the absence of muons and taus does not allow the
establishment of chemical equilibrium with the stellar plasma.
Therefore, they do not transport leptonic number outside the star
and their chemical potentials vanish. Their interactions with the
background are the same and no distinction needs to be made
between $\nu_{\mu}$ and $\nu_{\tau}$. Deeper in the protostar the
main interactions for these neutrinos are pair processes such as
$\mathcal{NN^{\prime
}}\rightleftarrows\mathcal{NN^{\prime}}\nu_{\mu,\tau}\bar{\nu}_{\mu,\tau
},e^{-}e^{+}\rightleftarrows\nu_{\mu,\tau}\bar{\nu}_{\mu,\tau}$,
etc, where $\mathcal{N}$ denotes a nucleon, and the scattering
reactions with nucleons and electrons. The scattering on electrons
are less frequent, but the energy transferred is greater. This
process dominates the energy exchange up to the energy sphere
where it freezes out. Beyond this radius, muon and tau neutrinos
still scatter off neutrons until they reach the neutrinosphere.
The cross section for
$\mathcal{N}\nu_{\mu,\tau}\rightleftarrows\mathcal{N}\nu_{\mu
,\tau}$ is smaller than the one of the $\beta$ reaction, and hence
the radius of the muon (tau) neutrinosphere $R_{\nu_{\mu,\tau}}$
is shorter than the radius of the electron neutrinosphere
$R_{\nu_{e}}$. The protoneutron star contracts, cools, and
deleptonizes due to the emission of neutrinos and antineutrinos
of all the flavors in a scale of time of the order of 10 sec. The
released energy is equally shared by all neutrino and
antineutrino species. This phase is known as the
Kelving-Helmholtz phase and ends up in the formation of a neutron
star.  During this time the radius of the protostar varies from
around $30$ to $10\,\mathrm{km}$, while the mass remains of the
order of $1.4\ M_{\odot}$. After approximately $1$ minute the
neutrino mean free path becomes comparable to the stellar radius
and neutrino luminosity decreases quickly.

Another important property predicted for neutron stars is that
they should have an extremely strong magnetic field. Since the
conductivity of the iron core is very large, the magnetic flux is
essentially trapped and it is conserved as it collapses to form a
neutron star\cite{meszaros}. Adopting for the initial magnetic
field the typical values observed in some white dwarfs, from the
flux conservation one can easily estimate that the surface field
of a protoneutron star has to be in the range of
$10^{12}-10^{13}\,\mathrm{G}$. This is supported by evidence
inferred from measurements of the spin-down rates and
observations of the X-ray spectra features\cite{urpin}. Magnetic
fields of the order of $10^{15}\,\mathrm{G}$ or stronger can be
generated by dynamo processes, and in the central regions
intensities up to $10^{18}\,\mathrm{G}$ have been
considered\cite{manka}.

\section{Momentum flux}

In a protoneutron star the momentum flux from the core to the
external regions is dominated by neutrinos, which provide the
most efficient mechanism to release the energy generated during
stellar collapse. As explained in the previous section, matter
densities are so high that neutrinos are trapped forming a
degenerated gas at quasi-equilibrium diffusing radially. The
statistical description of this gas is made in terms of a
distribution function $f_{\nu_{\ell}}(\mathbf{x},\mathbf{k},t)$.
If the action of external fields is negligible, then the neutrino
momentum between two consecutive collisions remains constant. The
change of the distribution function is due to neutrino collisions
with the particles in the medium and is described by the
Boltzmann equation
\begin{equation}
\frac{df_{\nu_{\ell}}}{dt}=C(f_{\nu_{\ell}})\,, \label{Boltz}%
\end{equation}
where $C(f_{\nu_{\ell}})$ is the collision integral and
\begin{equation}
\frac{df_{\nu_{\ell}}}{dt}=\frac{\partial f_{\nu_{\ell}}}{\partial
t}+\mathbf{\hat{k}}\cdot\nabla f_{\nu_{\ell}}\,.
\end{equation}

When the regime is close to equilibrium we can approximate the collision
integral as follows
\begin{equation}
C(f_{\nu_{\ell}})\simeq-\frac{f_{\nu_{\ell}}-f_{\nu_{\ell}}^{eq}}{\tau
_{\nu_{\ell}}}\,, \label{zz}%
\end{equation}
where $\tau_{\nu_{\ell}} \cong\lambda_{\nu_{\ell}}$ is the average time
between collisions and
\begin{equation}
f_{{\nu}_{\ell}}^{eq}(k)=\frac{1} {1+e^{(k-\mu_{\nu_{\ell}})/T}}%
\end{equation}
is the Fermi-Dirac distribution function for the gas of
relativistic neutrinos ($E_{\nu_{\ell}}\cong k=\left\vert
\mathbf{k}\right\vert $) with chemical potential
$\mu_{\nu_{\ell}}$ and temperature $T$. In the static situation
($\partial f_{\nu_{\ell}} / \partial t = 0$), Eq.~(\ref{Boltz})
with $C(f_{\nu_{\ell}}$) given by Eq.~(\ref{zz}) can be solved
iteratively. Proceeding in this way we obtain
\begin{equation}
f_{\nu_{\ell}}\simeq f_{\nu_{\ell}}^{eq}-\lambda_{\nu_{\ell}}\mathbf{\hat{k}}
\cdot\nabla f_{\nu_{\ell}}^{eq},
\end{equation}
which constitutes the diffusion approximation for the
distribution function. The second term in this expression gives
the correction to equilibrium and allows the transport of the
energy and momentum towards the exterior of the protostar.

Once the distribution function is known the neutrino energy-momentum tensor
can be computed as
\begin{equation}
T_{\nu_\ell}^{\mu\nu}=\int\frac{d^{3}k}{\left(  2\pi\right)
^{3}}\frac{k^{\mu }k^{\nu}}{k^{0}}f_{\nu_{l}}\,.
\end{equation}
From this formula, for the energy density we have
\begin{equation}
U_{\nu_{l}}\text{ }=T_{\nu_{l}}^{00}=\frac{1}{(2\pi)^{3}}\int k^{3}%
dk\sin\theta^{\prime}d\theta^{\prime}d\phi^{\prime}f_{\nu_{l}}=\frac
{F_{3}[\eta_{\nu_{l}}]}{2\pi^{2}}T^{4}\,, \label{u}%
\end{equation}
where $\eta_{\nu_{l}}=\mu_{\nu_{l}}/T$ is the degeneracy parameter and
$F_{3}[\eta_{\nu_{l}}]$ is a Fermi integral
\begin{equation}
F_{n}[\eta_{\nu_{l}}]=\int_{0}^{\infty}dxx^{n}[1+\exp(x-\eta_{\nu_{l}})]^{-1}.
\end{equation}
At each point within the protostar we choose a local frame
($x^{\prime },y^{\prime},z^{\prime}$) with
$\mathbf{\hat{r}}=\mathbf{\hat{z}}^{\prime}$,
such that $\mathbf{\hat{k}}=\cos\theta^{\prime}\mathbf{\hat{z}}^{\prime}%
+\sin\theta^{\prime}\cos\phi^{\prime}\mathbf{\hat{x}}^{\prime}+\sin
\theta^{\prime}\sin\phi^{\prime}\mathbf{\hat{y}}^{\prime}$.

The stress tensor can be expressed in terms of the energy density as
\begin{equation}
T_{\nu_{l}}^{ij}\text{ }=\delta^{ij}\frac{U_{\nu_{l}}}{3}. \label{ttt}%
\end{equation}
The equilibrium component of the distribution function is isotropic and gives
a null contribution to the momentum flux ($F_{\nu_{l}}^{i}=$ $T_{\nu_{l}}%
^{0i}$), which is generated exclusively by the diffusive term:
\begin{equation}
\mathbf{F}_{\nu_{\ell}}=\frac{1}{(2\pi)^{3}}\int k^{2}dk \sin\theta^{\prime
}d\theta^{\prime}d\phi^{\prime}\, \mathbf{k}f_{\nu_{\ell}}=-\frac{1}{6\pi^{2}%
}\int_{0}^{\infty}k^{3}\lambda_{\nu_{\ell}}\nabla f_{\nu_{\ell}}^{eq}dk\,.
\label{mflux}%
\end{equation}
To complete the calculation of $\mathbf{F}_{\nu_{\ell}}$ we need the mean free
paths of the neutrinos or equivalently, the opacities $\kappa_{\nu_{\ell}}$.

The neutral current reactions equally affect all flavors. Since
the energies available in the system are much smaller than the
mass of the $\tau$ lepton, the $\nu_{\tau}$ does not participate
in the charge-current interactions. For the same reason, and for
simplicity, we also ignore the presence of muons. Accordingly,
the main sources of neutrino opacities are
\begin{align}
\nu_{\ell}+n  &  \rightleftarrows\nu_{\ell}+n\,,\\
\nu_{\ell}+p  &  \rightleftarrows\nu_{\ell}+p\,,\\
\nu_{e}+n  &  \rightleftarrows e^{-}+p\,.
\end{align}
In terms of the cross sections for these reactions the $\lambda_{\nu_{\ell}%
}^{-1}$ are given by
\begin{align}
\lambda_{\nu_{e}}^{-1}  &  =\lambda_{\nu_{\mu,\tau}}^{-1}+N_{n}\sigma_{abs},\\
\lambda_{\nu_{\mu,\tau}}^{-1}  &  =N_{n}\sigma_{n}+N_{p}\sigma_{p}\,,
\end{align}
where $N_{n,p}$ are the neutron and proton number densities and
\begin{align}
\sigma_{n}  &  =\frac{G_{F}^{2}}{4\pi}\left(  1+3g_{A}^{2}\right)  k^{2}\,,\\
\sigma_{p}  &  =\sigma_{n}\left[  1-\frac{8\sin^{2}\theta_{W}}{1+3g_{A}^{2}%
}\left(  1-2\sin^{2}\theta_{W}\right)  \right]  \,,\\
\sigma_{abs}  &  =4\,\sigma_{n}\,. \label{abs}%
\end{align}
Here, $\sin^{2}\theta_{W}\simeq0.23$ and $g_{A}\simeq1.26$ is the
renormalization for the axial-vector current of the nucleons. In
Eq.~(\ref{abs}) we neglected contributions of order
$(m_{n}-m_{p})/k$, where $m_{n}$ and $m_{p}$ are the neutron and
proton mass, respectively.

From the previous formulae we immediately see that
\begin{equation}
\lambda_{\nu_{\ell}}^{-1}= \kappa_{\nu_{\ell}}\rho= \varkappa_{\nu_{\ell}%
}k^{2}\rho\,, \label{lll}%
\end{equation}
where the coefficients $\varkappa_{\nu_{\ell}}$ are constants. By making
$m_{n} \cong m_{p}$ and taking the typical values $Y_{n}\approx0.9$ and
$Y_{p}\approx0.1$ for the nucleon fractions $Y_{n,p}=N_{n,p}/(N_{n}+N_{p})$,
we get
\begin{align}
\varkappa_{\nu_{e}}  &  =3\times10^{-25}\,\mathrm{MeV}^{-5},\label{ce1}\\
\varkappa_{\nu_{\mu,\tau}}  &
=0.7\times10^{-25}\,\mathrm{MeV}^{-5}.\label{ce2}
\end{align}
When Eq.~(\ref{lll}) is substituted in the integrand of
Eq.~(\ref{mflux}), the momentum flux transported by neutrinos
diffusing from the core to the exterior of the protostar becomes
\begin{equation}
\mathbf{F}_{\nu_{\ell}}=-\frac{1}{6{\pi}^{2}}\frac{1}{ \varkappa_{\nu_{\ell}%
}\rho}\nabla\!\left(  F_{1}[\eta_{\nu_{\ell}}] \,T^{2}\right)  . \label{ff1}%
\end{equation}
In what follows we assume that the neutrino chemical potentials are
negligible, and using $F_{1}[0]=\pi^{2}/12$ and $F_{3} [0]=7\pi^{4}/120$, from
Eqs.~(\ref{u}) and (\ref{ff1}), we obtain
\begin{align}
U_{\nu_{\ell}}  &  =\frac{7\pi^{2}}{240}\,T^{4}\,,\label{u1}\\
\mathbf{F}_{\nu_{\ell}}  &  =-\frac{1}{72
\varkappa_{\nu_{\ell}}\rho}\,\nabla T^{2}\,.
\label{ff2}
\end{align}
The last results will be used in Section 6, when we calculate the
asymmetric neutrino emission in the context of a specific model
for the protoneutron star.

\section{Neutrino oscillations in a magnetized medium}

A notable consequence of neutrino mixing is the phenomena of
neutrino oscillations established five decades ago by
Pontecorvo\cite{pontecorvo, mohapal}. It corresponds to the
periodic variation in the flavor content of a neutrino beam as it
evolves from the production point. In the last years compelling
evidence of neutrino oscillations has been obtained in
experiments with atmospheric, solar, accelerator, and reactor
neutrinos\cite{exposc, exposc2}. The most plausible
interpretation for these new phenomena is that neutrinos are
massive particles and the weak-interaction states $\left\vert
\nu_{\ell}\right\rangle $ ($\ell =e,\mu,\tau$) are linear
combinations of the mass eigenstates $\left\vert
\nu_{j}\right\rangle$ ($j=1,2,3$), with masses $m_{j}$:
\begin{equation}
\left\vert \nu_{\ell}\right\rangle \mathbf{=\ }U_{\ell j}\ \left\vert \nu
_{j}\right\rangle , \label{mix}%
\end{equation}
where $U_{\ell j}$ are the elements of a unitary mixing matrix,
the analogue of the CKM matrix in the quark sector. Consider a
beam of $\nu_{\alpha}$ created at a certain point with a
3-momentum $\mathbf{k}$. When the state given in Eq.~(\ref{mix})
evolves, each of the mass eigenstates acquires a different phase
$e^{-iE_{j}t}$, with $E_{j} = \sqrt{m_{j}^{2} + k^{2}}$
($k=\left\vert \mathbf{k}\right\vert $). As a consequence, at a
certain distance $r\cong t$ from the source the neutrino beam
will not correspond to pure $\nu_{\ell}$ but turns partially into
other flavors.

The pattern of neutrino oscillations is modified when neutrinos
propagate through a material medium. The basic reason is that the
energy-momentum relation of an active neutrino is affected by its
coherent interactions with the particles that constitute the
medium. These modifications are described in terms of a potential
energy $V_{\ell}$, which is related to the real part of the
refraction index $n_{\ell}$ according to
$\mathrm{Re}\,n_{\ell}=1-V_{\ell}/k$ and can be calculated from
the background contributions to the neutrino
self-energy\cite{notraf}. If, like in normal matter, electrons
but not heavier leptons are present in the background, then only
$\nu_{e}$ scatter through charge current interactions, while all
the flavors interact via the neutral current. Therefore, the
refraction index differs for neutrinos of different flavors and
they will acquire distinct phases when propagating with the same
momentum through the medium. This fact can have impressive
consequences in the case of mixed neutrinos and, under favorable
conditions, flavor transformations are significantly enhanced in
a medium with a varying density even when mixing in vacuum is
small. This is the essence of the MSW mechanism\cite{msw,mohapal}
whose large mixing realization provides the solution to the solar
neutrino problem\cite{exposc2,solar}.

Besides the modifications of the dispersion relation, neutrinos
acquire in matter an effective coupling to the electromagnetic
field by means of their weak interaction with the background
particles\cite{DNEC}. As a result, in the presence of an external
magnetic field there are additional contributions to the neutrino
refraction index and flavor transformations are affected, but in
a way that preserves quirality\cite{oscB}, contrary to what
happens for neutrino oscillations driven by transition magnetic
moments\cite{oscmagm}. In what follows, we will discuss neutrino
oscillations in a magnetized medium for the simplest situation of
mixing between two neutrino species, that is, $\nu_{\ell}$ and
another active neutrino $\nu_{\ell^{\prime}}$ or a hypothetical
sterile neutrino $\nu_{s}$. In any case we can write
\begin{equation}
\mathbb{U}=\left(
\begin{array}
[c]{cc}%
\text{cos }\theta & \text{sin }\theta\\
-\text{sin }\theta & \text{cos }\theta
\end{array}
\right)  ,
\end{equation}
where $\theta$ is the vacuum mixing angle, a parameter to be fixed from the
experimental results.

The Dirac wave function for relativistic (mixed) neutrinos with
momentum $\mathbf{k}$ propagating in matter in the presence of a
static uniform magnetic field is well approximated as
follows\cite{oscB}:
\begin{equation}
\psi\cong e^{i\mathbf{k.x}}\left(
\begin{array}
[c]{c}%
0\\
\phi_{-}%
\end{array}
\right)  \chi(t)\, , \label{Diracwf}%
\end{equation}
where we use the Weyl representation, $\phi_{-}$ is the Pauli
spinor of negative helicity ($\vec{\sigma}.\hat{\varkappa}\
\phi_{-}=-\phi_{-}$) and $\chi(t)$ is a flavor-space spinor that
satisfies
\begin{equation}
i\frac{d\chi}{dt}=\mathbb{H}_{B}\chi\,. \label{wolfeq}%
\end{equation}
The Hamiltonian $\mathbb{H}_{B}$ is a $2\times2$ matrix that, in
the basis of the flavor states, takes the form
\begin{equation}
\mathbb{H}_{B}=k+\frac{\mathbb{M}^{2}}{2k}+\mathbb{V}\emph{,} \label{ham}%
\end{equation}
with
\begin{equation}
\mathbb{M}=\mathbb{U}\left(
\begin{array}
[c]{cc}%
m_{1} & 0\\
0 & m_{2}%
\end{array}
\right)  \mathbb{U}^{\dagger}%
\end{equation}
and $\mathbb{V}=\mathrm{diag}(V_{\ell},V_{\ell^{\prime},s})$.

The effective potential for active neutrinos can be written
as\cite{DNEC,sem}
\begin{equation}
V_{\ell}=b_{\ell}-c_{\ell}\,e\,\mathbf{\hat{k}}\cdot\mathbf{B}\,, \label{v2}%
\end{equation}
where $e>0$ is the proton charge. The coefficients $b_{\ell}$ and
$c_{\ell}$ depend on the properties of the thermal background. To
order $G_{F}$ the expression for $b_{\ell}$ is independent of
$B$, but this is not true in general for $c_{\ell}$. In our
analysis we will work within the weak-field limit, i.e.,
$B\ll\mu_{e}^{2}/2e$\cite{nu}, and ignore any possible dependence
of this coefficient on the magnetic field. The effect of strong
magnetic fields on the neutrino propagation in a medium have been
considered by some authors\cite{sem}. For (active) antineutrinos
the potentials change in sign, i.e.,
${\bar{V}}_{\ell}=-V_{\ell}$, while in the case of sterile
neutrinos, since they do not interact with the background, we have
\begin{equation}
V_{s}={\bar{V}}_{s}=0\,.\label{vs}%
\end{equation}

Eq.~(\ref{wolfeq}) is the extension of the MSW equation to the
situation we are considering, and was first derived in Ref.
\cite{oscB}. The components $\chi_{e}(t)$ and $\chi_{\mu,s}(t)$
of $\chi(t)$ give the amplitude to find the neutrino in the
corresponding flavor state. For a uniform background $\chi(t)$ =
$\sum\limits_{i=1,2}(\varsigma_{i}^{\dagger}\ \chi(0))\varsigma
_{i}e^{i\omega_{i}t}$, where $\omega_{i}(p)$ are the energies of
the neutrino modes in the medium and the vectors
$\varsigma_{1,2}$ are the solutions of the eigenvalue condition
$\mathbb{H}_{B}\varsigma_{i}=\omega_{i}\varsigma_{i}$. For
oscillations the relevant quantity is
\begin{equation}
\omega_{2}-\omega_{1}=\frac{1}{2k}\sqrt{\left( {\Delta m}^{2}\cos
2\theta-2k\mathcal{V}\right) ^{2}+\left( {\Delta m}^{2}\sin2\theta\right)
^{2}}\,,
\end{equation}
where $\mathcal{V}(r)=V_{\ell}(r)-V_{\ell^{\prime},s}(r)$ and
$\Delta m^{2}\equiv m_{2}^{2}-m_{1}^{2}$. When the medium is
inhomogenous, along the neutrino path ($r\cong t$) the elements
of $\mathbb{V}$ are functions of $t$, and $\chi(t)$ has to be
determined by solving Eq.~(\ref{wolfeq}) with $\mathbb{H}_{B}$ a
time dependent matrix. At each point $r$, $\mathbb{H}_{B}$ can be
diagonalized by the unitary transformation
\begin{equation}
\mathbb{U}_{m}(r)=\left(
\begin{array}
[c]{cc}%
\cos\theta_{m} & \sin\theta_{m}\\
-\sin\theta_{m} & \cos\theta_{m}%
\end{array}
\right) \,,\label{Umatter}%
\end{equation}
with the mixing angle in matter $\theta_{m}(r)$ given by
\begin{equation}
\tan2\theta_{m}=\frac{{\Delta m}^{2}\sin2\theta}{{\Delta m}^{2}\cos
2\theta-2k\mathcal{V}}\,.\label{eq.r43}%
\end{equation}
A similar expression is valid in the case of antineutrinos with
${\Delta m}^{2}\cos2\theta+2k\mathcal{V}$ in the denominator. In
general $\mathcal{V}$ and consequently the mixing angle in matter
depends on the magnetic field. As a function of $\mathcal{V}$,
$\sin^{2}2\theta_{m}$ exhibits the characteristic form of a
Breit-Wigner resonance. For neutrinos (antineutrinos) with
momentum $\mathbf{k}$ the resonance condition
($\sin^{2}2\theta_{m}=1$) is
\begin{equation}
\mathcal{V}(r_{_{R}})=\pm\,\frac{{\Delta
m}^{2}}{2k}\cos{2\theta}, \label{deltam}%
\end{equation}
where $r_{_{R}}= r_{_{R}}(\mathbf{k})$ and the sign $+$ ($-$)
corresponds to neutrinos (antineutrinos). Notice that for
active-active oscillations the neutral current contributions to
$\mathcal{V}=V_{\ell}-V_{\ell^{\prime}}$ cancel since they are
the same for both flavors. This is not true in the case of
active-sterile oscillations where $\mathcal{V}=V_{\ell}$, and
such contributions play a relevant role in the resonant
transformations for $\nu_{\ell}\longleftrightarrow\nu_{s}$ (or
$\bar{\nu}_{\ell}\longleftrightarrow\bar{\nu}_{s}$).The efficiency
of the flavor transformation depends on the adiabaticity of the
process. The conversion will be adiabatic when the oscillation
length at the resonance is smaller than the
width of the enhancement region. This imposes the condition%
\begin{equation}
\left\vert \frac{d\mathcal{V}}{dr}\right\vert_{r_{_{R}}}<\left(
\frac{\Delta m^{2}}{2k}\sin2\theta\right)^{\!2}. \label{adcon}%
\end{equation}

The dispersion relation of neutrinos in a magnetized plasma made
of electrons, protons, and nucleons have been calculated in
Ref.~\cite{DN}, including the contributions due to the anomalous
magnetic moment coupling of the nucleons to the photon. From the
results given there we have
\begin{align}
b_{e}  &  =\sqrt{2}G_{F}\left[  N_{e}+\left(  \frac{1}{2}-2\sin^{2}\theta
_{W}\right)  (N_{p}-N_{e})-\frac{1}{2}N_{n}\right]  +\tilde{b}_{_{e}},\\
b_{\mu,\tau}  &  =\sqrt{2}G_{F}\left[  \left(  \frac{1}{2}-2\sin^{2}\theta
_{W}\right)  \left(  N_{p}-N_{e}\right)  -\frac{1}{2}N_{n}\right]  +\tilde
{b}_{_{_{\mu,\tau}}},\\
c_{e}  &  =2\sqrt{2}G_{F}\left[  g_{A}\left( 1+\,2m_{p}\frac{\kappa_{p}}%
{e}\right)  C_{p}-g_{A}2m_{n}\frac{\kappa_{n}}{e}C_{n}-C_{e}\right]  ,\\
c_{\mu,\tau}  &  =2\sqrt{2}G_{F}\left[ g_{A}\left(  1+\,2m_{p}\frac{\kappa
_{p}}{e}\right)  C_{p}-g_{A}2m_{n}\frac{\kappa_{n}}{e}C_{n}+C_{e}\right]  ,
\end{align}
where $\kappa_{n,p}$ are the anomalous part of the nucleon magnetic moments
given by
\begin{equation}
\kappa_{n}=-1.91\frac{e}{2m_{n}} \ ,
\ \ \ \ \ \
\kappa_{p}=1.79\frac{e}{2m_{p}} \ ,
\end{equation}
and
\begin{align}
\tilde{b}_{\ell}  &  =\sqrt{2}G_{F}\left[  N_{\nu_{\ell}}-N_{\bar{\nu}_{\ell}%
}+\sum_{_{\ell^{\prime}}}\left(  N_{\nu_{\ell^{\prime}}}-N_{\bar{\nu}%
_{\ell^{\prime}}}\right)  \right]  \,,\\
C_{\flat}  &  =\frac{1}{2}\int\frac{d^{3}p}{(2\pi)^{3}}\frac{1}{2E}\frac
{d}{dE}f_{\flat}\ \ \ \ (\flat=e,n,p)\,.
\end{align}
In these equations $N_{e}$ and $N_{\nu_{\ell}}$($N_{\bar{\nu}_{\ell}}$) are
the number densities of electrons and neutrinos (antineutrinos) of flavor
$\ell$, respectively and $f_{\flat}$ are the distribution functions of the
electrons, protons, and neutrons.

We model the atmosphere of a protoneutron star as a neutrino gas
diffusing through a plasma made of relativistic degenerate
electrons and classical nonrelativistic nucleons. In such
conditions $C_{e}=-\mu_{e}/8\pi^{2}$ and
$C_{n,p}=-N_{n,p}/8Tm_{n,p}$\cite{DN,nu}, where $T$ is the
background temperature and $\mu_{e}=\left(  3\pi^{2}N_{e}\right)
^{1/3}$ is the electron chemical potential. The neutrino
fractions are much smaller than the electron fraction
$Y_{e}=N_{e}/(N_{n}+N_{p})$ $\sim0.1$ and, in the context of our
analysis, the contributions to $b_{\nu_{\ell}}$ coming from the
neutrino-neutrino interactions (denoted by $\tilde{b}_{_{\ell}}$)
can be neglected. With this approximation in mind, putting
$m_{p}\simeq m_{n}$, and taking into account that by electrical
neutrality $N_{e}=$ $N_{p}$, the coefficients in the effective
potentials become:
\begin{align}
b_{e}  &  \simeq-\frac{G_{F}}{\sqrt{2}}\,\left(  1-3Y_{e}\right)  \frac{\rho
}{m_{n}} \ ,\label{bn}\\
b_{\mu,\tau}  &  \simeq-\frac{G_{F}}{\sqrt{2}}\,\left(  1-Y_{e}\right)
\frac{\rho}{m_{n}} \ , \label{bn2}\\
c_{e}  &  \simeq-\sqrt{2}\,G_{F}\left(  \frac{3+Y_{e}}{5Tm_{n}^{2}}%
\,\rho-\frac{\mu_{e}}{4\pi^{2}}\right) , \label{cn2}\\
c_{\mu,\tau}  &  \simeq-\sqrt{2}\,G_{F}\left(  \frac{3+Y_{e}}{5Tm_{n}^{2}%
}\,\rho+\frac{\mu_{e}}{4\pi^{2}}\right) . \label{cn}%
\end{align}
It should be noticed that in Ref.~\cite{KSS} $c_{e}$ vanishes because all the
components of the stellar plasma are assumed to be degenerated, and the
contributions due to the anomalous magnetic moment of the nucleons are not included.

We now use the effective potentials $V_{\ell}$ with $b_{\ell}$ and
$c_{\ell}$ given by Eqs.~(\ref{bn})-(\ref{cn}) to examine the
magnetic field effect on the resonant neutrino oscillations
within a protoneutron star. In a linear approximation, we can
write
\begin{equation}
r_{_{R}}(\mathbf{k})=r_{o}(k)+\,\delta(k)\cos\alpha\,, \label{rR}
\end{equation}
where $r_{o}$ and $\delta$ are ordinary functions of the
magnitude of the neutrino momentum and $\alpha$ is the angle
between $\textbf{k}$ and $\textbf{B}$. For a certain value of $k$,
Eq.~(\ref{rR}) determines a spherical shell limited by the
spheres of radii $r_{o}(k)\pm\,\delta(k)$, where the resonance
condition is verified for neutrinos (or antineutrinos) moving in
different directions with respect to the magnetic field. The
quantity $r_{o}$ corresponds to the radius of the resonance
sphere when $B=0$ and in the case of oscillations between
$\nu_{e}$ and $\nu_{\mu}$ (or $\nu_{\tau}$) is given by
\begin{equation}
\left.  G_{F}\sqrt{2}\,Y_{e}\frac{\rho}{m_{n}}\right\vert
_{r_{o}}=\pm
\frac{\Delta m^{2}}{2k}\cos{2\theta}\,. \label{recon}%
\end{equation}
The left-hand side corresponds to the difference
$b_e-\,b_{\mu,\tau}$ evaluated at $r=r_{o}$ and is positive
definite. Therefore, assuming $\Delta m^{2}\cos2\theta>0$, the
above condition can be satisfied for $\nu
_{e}\longleftrightarrow\nu_{\mu,\tau}$ but not for $\bar{\nu}_{e}%
\longleftrightarrow\bar{\nu}_{\mu,\tau}$. To be effective as a
kick mechanism the resonance transformations of active neutrinos
have to take place in the region between the electron and muon
(tau) neutrinospheres. In this case, the $\nu_{e}$ are trapped by
the medium, but the $\nu_{\mu,\tau}$ produced this way are above
their neutrinosphere and, when moving outside, can freely escape
from the star. In the presence of a magnetic field, the emission
points for $\nu_{\mu,\tau}$ having the same $k$ but different
directions are not at the same radius, which originates an
asymmetry in the momentum they carried away.

Replacing $k$ in Eq.~(\ref{recon}) by the thermal average $\left\langle
k\right\rangle \simeq3.15\,T$ and taking into account that $\rho\sim
10^{11}\,\mathrm{g}\,\mathrm{cm}^{3}$ and $T\sim4\mathrm{\,MeV}$ at
$R_{\nu_{e}}$, the requirement of having the resonance within the electron
neutrinosphere implies $\Delta m^{2}\cos{2\theta}\gtrsim2\times10^{4}%
\,\mathrm{eV}^{2}$, which is excluded by the experimental results
on neutrino oscillations and the cosmological limits. For this
reason, in what follows we concentrate only on the active-sterile
oscillations. In this case, the resonance condition when $B=0$
takes the form
\begin{equation}
b_{\ell}(r_{o_{\ell}})=\mp\frac{\Delta m^{2}}%
{2k}\cos{2\theta}\,,\label{recons1}
\end{equation}
for $\nu_{\ell}\longleftrightarrow\nu_{s}$ ($\bar{\nu}_{\ell}
\longleftrightarrow\bar{\nu}_{s}$). In the static model we use for
the protoneutron star, a good approximation is to take
$Y_{e}<1/3$. Therefore, if $\Delta m^{2}\cos2\theta>0$, the
resonant condition is verified only by antineutrinos and from now
on we restrict ourselves to this situation.

To determine the quantity $\delta$ in Eq.~(\ref{rR}) we substitute
$V_{\ell}$ as given by Eq.~(\ref{v2}) in Eq.~(\ref{deltam}) and
expand $b_{\ell}$ up to first order in $\delta_{\ell}$.
Proceeding in this way we get
\begin{equation}
\delta_{\ell}(k)=\mathcal{D}_{\ell}(k)\,eB\,,
\end{equation}
with
\begin{equation}
\mathcal{D}_{\ell}(k)=\left.
\frac{1}{h_{b_{\ell}}^{-1}}\frac{c_{\ell}}{b_{\ell} }\right\vert
_{r_{o_{\ell}}}, \label{d}
\end{equation}
where we have defined $h_{g}^{-1}\equiv\frac{d}{dr}\ln g$ for any
function $g(r)$. Using Eqs.~(\ref{bn})-(\ref{cn}) we find
explicitly
\begin{align}
\mathcal{D}_{e} &  =\left.  \frac{2}{\left(  1-3Y_{e}\right)  h_{\rho}^{-1}%
-3Y_{e}h_{Y_{e}}^{-1}}\left(  \frac{3+Y_{e}}{5Tm_{n}}-\frac{m_{n}\mu_{e}}%
{4\pi^{2}\rho}\right)  \right\vert _{r_{o_{e}}},\label{d3}\\
\mathcal{D}_{\mu,\tau} &  =\left.  \frac{2}{\left(  1-Y_{e}\right)  h_{\rho}%
^{-1}-Y_{e}h_{Y_{e}}^{-1}}\left(  \frac{3+Y_{e}}{5Tm_{n}}+\frac{m_{n}\mu_{e}%
}{4\pi^{2}\rho}\right)  \right\vert _{r_{o_{\mu,\tau}}}.\label{d2}%
\end{align}

Finally from Eq.~(\ref{adcon}), discarding terms proportional to
the magnetic field, the following restriction to the vacuum
mixing angle results
\begin{equation}
\tan^{2}2\theta>\left\vert
\frac{h_{b_{\ell}}^{^{{-1}}}}{b{_{\ell}}}
\right\vert_{r_{o_{\ell}}}, \label{adcon2}
\end{equation}
which guarantees that the evolution will be adiabatic and an
almost complete flavor transformation for small values of
$\theta$.

\section{Neutrino momentum asymmetry}

If the temperature and density profiles are isotropic, then the
momentum flux in the protoneutron star atmosphere is radial (see
Eq.~(\ref{ff2})). As a consequence, in most of the works on the
kick mechanism driven by matter neutrino oscillations the
resonance condition was evaluated assuming that neutrinos move
outside with a momentum $\left\langle k\right\rangle = (7\pi^{4}
/180\,\zeta(3))\,T\simeq 3.15\,T$ pointing in the radial
direction. In presence of a magnetic field this approach leads to
the concept of a deformed resonance surface, which acts as an
effective neutrinosphere. However, at each point within the
neutrinosphere there are neutrinos with momentum $\mathbf{k}$
pointing in every directions and the resonance condition as given
by Eq.~(\ref{deltam}) actually defines a spherical surface for
each value of $\cos\alpha$. This surface acts as a source of
sterile neutrinos produced through the resonant
$\bar{\nu}_{\ell}$ transformations, which move in the
$\mathbf{\hat{k}}$ direction. Therefore, as was mentioned above,
for a given $k$ the resonant transformations take place within a
spherical shell, and not on a deformed spherical surface. A more
careful calculation of the pulsar kick along this line has been
carried out in Ref.~\cite{n3} and here we follow the same
approach.

Let us consider the situation where the resonance shell is totally
contained within the $\bar{\nu}_{\ell}$ neutrinosphere. The
sterile neutrinos produced there that move out freely escape from
the protostar, but those directed toward the inside cross the
resonance surface again and reconvert into $\bar{\nu}_{\ell}$
which, being within their own neutrinosphere, are thermalized.
Consequently, only those $\bar{\nu}_{s}$ going outward can leave
the star and the resonance surfaces for different momentum
directions behave as effective emission semispheres of sterile
antineutrinos. For $\mathbf{k}$ pointing in opposite directions
the radii of the respective semispheres are
$r^{\pm}_{_R}=r_{o}(k)\pm\,\delta(k)\left\vert
\cos\alpha\right\vert$, and therefore they have different areas
that generate a difference in the momentum carried away by the
sterile neutrinos leaving the star in opposite directions. There
also exists a temperature variation within the resonance shell
such that the temperatures of the emission semispheres for
$\bar{\nu}_{s}$ going outside in opposite directions are different
($T(r^{-}_{_R})>T(r^{+}_{_R})$). The variations in the area and
the temperature tend to compensate each other when the
contributions of $\bar{\nu}_{s}$ emitted in every direction are
added up. In the calculation below we take both semispheres into
account when explicitly computing the asymmetry in the total
momentum $\mathcal{K}$ emitted by the cooling protoneutron star.
In this way, we improve the calculation of Ref.~\cite{n3} where
the temperature within the resonance shell was assumed to be
uniform.

The active neutrinos and antineutrinos are emitted isotropically
from their respective neutrinospheres and do not contribute to the
momentum asymmetry $\Delta\mathcal{K}$. Therefore, the only
nonvanishing contribution can come from the $\bar{\nu}_{s}$. If
their emission lasts for an interval $\Delta t$ of the order of a
few seconds, then in the static model we are using for the
protostar $\Delta \mathcal{K}= \left\vert K_{_B} \right\vert\Delta
t$, where $K_{_B}$ is the component along the magnetic field
direction of the neutrino momentum emitted per time unit.
According to Eq.~(\ref{mflux}) we can write
\begin{equation}
K_{_B}=
\frac{1}{(2\pi)^{2}}\!\int_{0}^{\infty}k^{3}dk\int_{0}^{\pi}\sin\theta
d\theta\int_{0}^{\frac{\pi}{2}}\sin\theta^{\prime}
d\theta^{\prime}\int_{0}^{2\pi}d\phi^{\prime}\,{r^2_{_R}}_{_\ell}(\mathbf{k})
f_{_{\bar{\nu}_s}}\mathbf{\hat{k}\cdot\hat{B}}\,, \label{kB}
\end{equation}
where $\mathbf{\hat{k}\cdot\hat{B}}=
\cos\theta\cos\theta^{\prime}-\sin\theta
\sin\theta^{\prime}\sin\phi^{\prime}$ and the distribution
function for sterile neutrinos $f_{_{\bar{\nu}_s}}$ is evaluated
at $r_{_{R_{\ell}}}$. We have chosen a reference frame ($x$, $y$,
$z$) fixed to the protostar where the magnetic field coincides
with the $z$-axis, $\mathbf{B}=B\,\mathbf{\hat{z}}$. In addition,
at each point within the resonance shell we use the local frame
introduced in Section 2 to evaluate the components of the
energy-momentum tensor. Here, the upper limit for the integration
on $\theta^{\prime}$ is $\pi/2$ instead of $\pi$, because we have
to include only the contributions from the sterile neutrinos going
outside.

To proceed further with the calculation of $K_{_B}$ we need to
know $f_{_{\bar{\nu}_s}}$. The proper treatment of the interplay
of collisions and oscillations for trapped neutrinos requires the
use of the density matrix. For our purpose a less elaborate
description will suffice and we put $f_{_{\bar{\nu}_{s}}}\!\cong
\mathcal{P}(\bar{\nu}_{\ell}\rightarrow\bar{\nu}_s)f_{_{\bar{\nu}_{\ell}}}$,
where $\mathcal{P}(\bar{\nu}_{\ell}\rightarrow\bar{\nu}_s)$ is the
probability for $\bar{\nu}_{\ell}$ to convert into $\bar{\nu}_s$.
Let us assume that the adiabaticity condition given by
Eq.~(\ref{adcon2}) is verified in the interior of the
$\bar{\nu}_{\ell}$ neutrinosphere, and hence for small mixing
$\mathcal{P}(\bar{\nu}_{\ell}\rightarrow\bar{\nu}_{s})\cong 1$.
Accordingly in Eq.~(\ref{kB}) we take
$f_{_{\bar{\nu}_{s}}}(r_{_R})\simeq\,
f_{_{\bar{\nu}_{\ell}}}(r_{_R})$, with
\begin{equation}
f_{_{\bar{\nu}_{\ell}}}(r_{_R})\cong f_{_{\bar{\nu}_{\ell}}}^{eq}
(r_{o_\ell})-\!\left.\left(
\frac{\cos\theta^{\prime}}{\varkappa_{_{\bar{\nu}_{\ell}}}k^{2}\rho}
\frac{df_{_{_{\bar{\nu}_{\ell}}}}^{eq}}{dr}-\mathbf{\hat{
k}\cdot\hat{B}}\,\frac{df_{_{_{\bar{\nu }_{\ell}}}}^{eq}}{dr}\,
\delta_{\ell}(k)\right)\right\vert_{r_{o_\ell}}, \label{expf}
\end{equation}
where a term proportional to the derivative of
${\rho}^{-1}df_{_{\bar{\nu}_{\ell}}}^{eq}/dr$ has been neglected,
in agreement with the diffusive approximation we are employing.
Substituting the expansion (\ref{expf}) into Eq.~(\ref{kB}) and
keeping terms that  are at most linear in $\delta_{\ell}$ we find
\begin{align}
K_{_B} \cong & \,\frac{2}{3\pi}\!\int_{0}^{\infty}dk\
k^{3}r_{o_{\ell}}(k)\,\delta_{\ell}(k)\nonumber\\
& \left.\!\left(f_{_{_{\bar{\nu}_{\ell}}}}^{eq}
-\frac{1}{2\varkappa_{_{\bar{\nu}_{\ell}}
}k^{2}\rho}\frac{df_{_{_{\bar{\nu}_{\ell}}}}^{eq}}{dr} +
\frac{1}{2}r_{o_{\ell}}(k)\,
\frac{df_{_{_{\bar{\nu}_{\ell}}}}^{eq}}{dr}\,\delta_\ell(k)\right)\right\vert
_{r_{o_{\ell}}}. \label{eb}
\end{align}
The energy released during the core collapse of the progenitor
star is approximately equipartitioned among all the neutrino and
antineutrino flavors. In the context of our analysis this means
that $K_{r}\,\Delta t\simeq \mathcal{K}/6$, where $K_r$ represents
the momentum rate of the sterile neutrinos in the radial direction
and is given by an expression similar to Eq.~(\ref{kB}) with
$\mathbf{\hat{k}\cdot\hat{B}}$ replaced by
$\mathbf{\hat{k}\cdot\hat{r}}$. Following the same procedure
outlined above we find
\begin{equation}
K_{r} \cong \frac{1}{2\pi}\!\int_{0}^{\infty}dk\ k^{3}\
r_{o_{\ell}}^{2}(k)\left.\!\left(
f_{_{_{\bar{\nu}_{\ell}}}}^{eq}-\frac{2}{3\varkappa_{_{\bar{\nu}_{\ell}}}\rho
k^{2}}\frac{df_{_{_{\bar{\nu}_{\ell}}}}^{eq}}{dr}\right)
\right\vert _{r_{o_\ell}}\,.\label{er}
\end{equation}

In terms of $K_{_B}$ and $K_r$ the fractional asymmetry in the
total momentum neutrinos carry away becomes
\begin{equation}
\frac{\Delta\mathcal{K}}{\mathcal{K}}=\frac{\left\vert
K_{B}\right\vert}{6K_{r}}\,. \label{asym}
\end{equation}
To evaluate the remaining integrals in Eqs.~(\ref{er}) and
(\ref{eb}) the explicit dependence on $k$ of the functions
$r_{o_\ell}$ and $\mathcal{D}_\ell$ have to be known. Simple
analytical results can be derived by replacing these functions by
the constant quantity $\bar{r}_{o_\ell}$ and
$\overline{\mathcal{D}}_\ell=\mathcal{D}_\ell(\bar{r}_{o_\ell})$,
where $\bar{r}_{o_\ell}$ is determined from the resonance
condition when $k$ is replaced by its thermal average
$3.15\,T(r)$:
\begin{equation}
6.3\,T(\bar{r}_{o_\ell})\,b_{\ell}(\bar{r}_{o_\ell})=\,\Delta
m^{2}\cos2\theta\,.
\end{equation}
Proceeding in this manner we get
\begin{align}
K_{r}  & \cong
\frac{\pi\bar{r}_{o_\ell}^{2}}{48}\left.\left(\frac{7\pi^{2}T^{4}}{5}
-\frac{4}{3\kappa_{_{\bar{\nu}_{\ell}}}\rho}\frac{dT^{2}}{dr}\right)\right\vert
_{\bar{r}_{o_\ell}}
,\label{erad}\\
K_{B}  & \cong
\frac{\pi\bar{r}_{o_\ell}\bar{\delta_\ell}}{36}\left.\left(\frac{7\pi^{2}T^{4}}{5}+
\frac{7\pi^{2}\bar{r}_{o_\ell}}{10}\frac{dT^{4}}{dr}-\frac{1}
{\varkappa_{_{\bar{\nu}_{\ell}}}\rho}\frac{dT^{2}}{dr}\right)\right\vert
_{\bar{r}_{o_\ell}}. \label{brad}
\end{align}
In these expressions the terms that depend on $dT^{2}/dr$ come
from the diffusive part of the distribution function. In the
regime considered here they are smaller than the other
contributions and in what follows we will neglect them.

Suppose that the pulsar momentum $K_{\!pul}$ is entirely due to
the active-sterile oscillation driven kick, then
$\Delta\mathcal{K}= K_{\!pul}$. For a pulsar with a mass
$M\!\approx M_{\odot}$ and a translational velocity
$v\approx500\,\rm{km}\,\rm{s}^{-1}$ we have
$K_{\!pul}\!\approx10^{41}\,\rm{g}\,\rm{cm}\,\rm{s}^{-1}$. Since
neutrinos are relativistic, the total amount of momentum they
carry is equal to the gravitational energy liberated by means of
them, that is
$\mathcal{K}\approx10^{43}\,\rm{g}\,\rm{cm}\,\rm{s}^{-1}$.
Therefore, as mentioned in the Introduction,
$\Delta\mathcal{K}/\mathcal{K}$ must be of the order of 0.01 to
obtain the required kick. In this case Eq.~(\ref{asym}) yields
\begin{equation}
eB \cong \frac{0.045\,\bar{r}_{o_{\ell}}}{\left\vert
(1+2\bar{r}_{o_{\ell}}
\bar{h}_{T}^{-1})\,{\overline{\mathcal{D}}_{\ell}}\right\vert}\,,
\label{momas}
\end{equation}
with $\bar{h}_{T}^{-1}$ denoting the logarithmic derivative of
$T(r)$ evaluated at $\bar{r}_{o_{\ell}}$. This formula allows us
to determine the magnitude of the magnetic field once the
temperature, the baryon density, and the electron fraction
profiles are known. Below we do it for the
$\bar{\nu}_{\mu}\!-\!\bar{\nu}_s$ conversion under the assumption
that the resonance layer is entirely within the
$\bar{\nu}_{\mu}$-sphere. However, we will first indicate how
$Y_e(r)$ is determined by means of $T(r)$ and $\rho\,(r)$ within
our approximate description of a protoneutron star.

For $\mu_{_{\nu _{e}}}\!\simeq 0$, the $\beta$ equilibrium gives
the relation $\mu _n-\mu _p\simeq\mu _e$ among the chemical
potentials of the background fermions. Hence, for
non-relativistic nucleons $N_{n}/N_{p}\simeq\,\exp[-(m_n-m_p-\mu
_{e})/T]$ and
\begin{equation}
Y_{e}\simeq \frac{1}{1+e^{\mu _{e}/T}}\,,
\label{ye}
\end{equation}
where we used the fact that in the condition prevailing in a
protoneutron star $\mu_e$ is much larger than the difference
between the neutron and the proton masses. Since electrons are
degenerated $\mu_e=\left(3{\pi}^2 \rho Y_e/m_n \right)^{\!1/3}$\,
and Eq. (\ref{ye}) is an implicit equation for $Y_{e}$ to be
solved numerically once $\rho$ and $T$ are known as a function of
$r$. Taking the derivative of Eq.~(\ref{ye}) with respect to $r$,
$h_{Y_{e}}^{-1}$ can be expressed in terms of $h_{\rho }^{-1}$ and
$h_{T}^{-1}$ as follows
\begin{equation}
h_{Y_{e}}^{-1}=-\left( h_{\rho }^{-1}-3h_{T}^{-1}\right)
\frac{1-Y_{e}}{1-Y_{e}+\frac{3T}{\mu _{e}}}\,.  \label{hy}
\end{equation}

To estimate $B$ we now assume that the density and the temperature
are related according to
$\rho=\rho_{c}\left(T/\,T_{c}\right)^{3}$\cite{BM}, where
$\rho_{c}$ and $T_{c}$ are the respective values of these
quantities at the core radius $R_{c}$. From this expression we
see that $h_{\rho}^{-1}=\,3\,h_{T}^{-1}$ which implies that
$h_{Y_{e}}^{-1}=0$ (see Eq.~(\ref{hy})). Therefore, the electron
fraction $Y_{e}$ and as a consequence the ratio $\mu_{e}/\,T$ are
constants. As core parameters we take $R_{c}=10\,\textrm{km}$,
$\rho _{c}=10^{14}\,\rm{g}\,\rm{cm}^{-3}$, and
$T_{c}=40\,\rm{MeV}$, while for the density profile we adopt the
potential law $\rho\,(r)=\rho_{c}\,(R_{c}/r)^{4}$ which yields
$h_{\rho}^{-1}= -4/r$ and $T(r)=T_{c}\,(R_{c}/r)^{4/3}$. The
$\bar{\nu }_{\mu}$-sphere is taken stationary at a density of
about $10^{12}\,\rm{g}\,\rm{cm}^{-3}$ that corresponds to a radius
$R_{\bar{\nu}_{\mu}}\simeq3\,R_{c}$ and a temperature
$T_{_{R_{\bar{\nu}_{\mu}}}}\simeq9\,\rm{MeV}$. Using these results
as well as the formula for $\mathcal{D}_\mu$ given in
Eq.~(\ref{d2}), from Eq.~(\ref{momas}) we obtain the following
simple result:
\begin{equation}
B=\,5\times10^{17}\left(1+0.3\frac{T_{c}}{\bar{T}}\right)
^{\!-1}\frac{\bar{T}}{T_{c}}\,\rm{G}\,,
\end{equation}
where $\bar{T}=T(\bar{r}_{o_{\mu}})$, with $R_c\leq
\bar{r}_{o_{\mu}}\leq\,R_{\bar{\nu}_{\mu}}$. We see that the
required intensity of the magnetic field decreases monotonically
from $3.8\times10^{17}\,\rm{G}$ at the core radius to
$4.8\times10^{16}\,\rm{G}$ at the $\bar{\nu}_{\mu}$-sphere.
According to Eqs.~(\ref{recons1}) and (\ref{bn2}), within the
specified interval for $\bar{r}_{o_{\mu}}$ the range of the
allowed values of the oscillation parameters are $7\times
10^{8}\,\rm{eV}^{2}\gtrsim \Delta m^{2}\cos 2\theta
\gtrsim 10^{5}\,\rm{eV}^{2}$, and for small mixing we have $25\,\rm{KeV}\gtrsim $ $%
m_{s}\gtrsim 0.3\,\rm{KeV}$. The requirement that the resonant
conversion be adiabatic imposes the condition
$\tan^{2}2\theta\gtrsim3\times10^{-11}\left(\bar{r}_{o_{\mu}}/R_{c}\right)
^{3}$ on the mixing angle. A last comment is in order, the ratio
of the diffusive term to the isotropic one in Eqs.~(\ref{erad})
and (\ref{brad}) can be expressed as
$10^{-4}\,(\bar{r}_{o_{\mu}}/R_{c})^{17/3}$, which illustrates the
validity of the approximation done in writing Eq.~(\ref{momas}).
In the next section we repeat these calculations in the realm of
the spherical Eddington model, which provides a simple but quite
physically reasonable description of a static neutrino atmosphere.

\section{Spherical Eddington Model}

The actual configuration of the neutrinosphere and the resonance
region depend on details of the protoneutron star, such as the
density, temperature, pressure, and leptonic fraction profiles.
The asymmetry in the momentum flux also depends on these features
and on the configuration of the magnetic field. The present
knowledge about protoneutron stars is not enough to single out a
definite model for their structure, but there are some well
established general characteristics that help towards proposing
models that are both workable and plausible. An important feature
is that the size and the shape of the different profiles do not
suffer sudden changes during the time interval when the main
neutrino emission takes place, i.e. $0.5\,s\lesssim
t\lesssim10\,s$. Thus, as a first approximation we can consider a
stationary regime that greatly simplifies calculations. Another
characteristic to be taken into account is that the bulk of
neutrinos is produced in the inner core. Consequently, we assume
that there is no neutrino production outside the core and hence
the flux of diffusing neutrinos is conserved in this region. This
flux is responsible for the transport of the energy liberated by
the protostar.

As mentioned before, the system is supposed to be constituted by
nucleons, with a density $\rho$, electrons, neutrinos and
antineutrinos. Neutrinos are in thermal equilibrium with nucleons,
satisfying a transport regime consistent with the diffusion
approximation, and are assumed to have a vanishing chemical
potential, $\mu_{\nu}=\mu_{\bar{\nu} }=0$. The total energy
density, the radial energy flux, and the pressure for neutrinos
and antineutrinos in the case of spherical symmetry, can be
obtained from Eqs. (\ref{u1}), (\ref{ff2}) and (\ref{ttt})%
\begin{align}
U &  =\underset{\ell}\sum\,U_{\nu_\ell}=
\frac{7}{40}\pi^{2}T^{4},\label{t1}\\
F & =\underset{\ell}\sum\,F_{\nu_{\ell}}=
-\frac{1}{36\bar{\varkappa}\rho}\ \frac{dT^{2}}{dr}\ ,\label{t2}\\
P_{\nu} & =\underset{\ell}\sum\,T_{\nu_{\ell}}^{ii}=\frac{U}
{3}\,,\label{t3}
\end{align}
where ${\bar{\varkappa}}^{-1}=\varkappa_{_{\nu_{e}}}
^{-1}+2\varkappa_{_{\nu_{\mu,\tau}}}^{-1}$. The main contributions
to the neutrino opacities were discussed in Section III. Using the
results given there for $\varkappa_{_{\nu_{e}}}$ and
$\varkappa_{_{\nu_{\mu,\tau}}}$ (Eqs.~(\ref{ce1}) and (\ref{ce2}))
we obtain $\bar{\varkappa}=3\times10^{-26} \ \rm{MeV}^{-5}$.

If in addition we assume that baryons constitute a nonrelativistic
ideal gas, we have the Eddington model. It is simple and
physically well justified, allowing a detailed discussion of the
relevant characteristics of the protostar and the geometry of the
resonance region. Originally proposed to describe a stellar
photosphere, this model was adapted by Schinder and
Shapiro\cite{ss} to a neutrino atmosphere with a plane geometry.
The more realistic case of a spherical atmosphere was considered
in Ref.~\cite{BDMZ} to analyze the geometrical effect on the
asymmetric neutrino emission induced by resonant oscillations.
Photons and electrons are of course present and, in fact,
electrons make the leading contribution to the effective potential
in the case of oscillations among active neutrinos. However, we
can ignore both photons and electrons for the hydrodynamical
description of the system.

The state equation that defines the system is given by the sum of
the contributions from nucleons and neutrinos to the pressure gas,
which is
\begin{equation}
P=\frac{\rho T}{m_{n}}+\frac{U}{3}. \label{eqst}
\end{equation}
In the Newtonian limit for the metric ($g_{oo} = -1 - 2\phi$,
$g_{oj}= 0$, and $g_{ij}= \delta_{ij}$, with
$\phi=-\,{GM(r)}/{r}$) the conservation of the total
(neutrinos\,+\,matter) energy-momentum tensor yields the energy
flux conservation
\begin{equation}
\frac{\partial (r^2F)}{\partial r}=0 \label{consf}
\end{equation}
and the hydrostatic equilibrium equation
\begin{equation}
\frac{dP}{dr}=-\left(P+\rho+U\right)\frac{GM}{r^{2}}\,.
\label{gradp}
\end{equation}
Here,
$M(r)=4\pi\int_{0}^{r}dr^{\prime}\,r^{\prime2}\rho\,(r^{\prime})$
is the mass enclosed up to a distance $r$ from the center.

Through the atmosphere the baryon density is
$\rho\simeq(10^{11}-10^{14})\,\rm{g}\,\rm{cm}^{-3}$ and
$T\simeq(4-40)\,\rm{MeV}$, and thus we have $U\simeq
(10^{-3}-10^{-2})\,\rho$ and $P\simeq (4\times10^{-3}
-4\times10^{-2})\,\rho$. This means that on the right-hand side of
Eqs.~(\ref{eqst}) and (\ref{gradp}) we can ignore the
contributions coming from the energy density of neutrinos and the
baryonic density. In addition, Eq.~(\ref{consf}) implies that $F=
L_{c}/{4\pi r^{2}}$, where $L_{c}$ is the luminosity of the
protostar. Taking all the above into account we arrive at the
following set of equations for an isotropic atmosphere of
neutrinos in thermal equilibrium with an ideal gas of nucleons:
\begin{align}
\frac{dT^{2}}{dr} & = -\rho\,\frac{9\bar{\varkappa}\,L_{c}}{\pi
r^{2}}
\,,\label{m2}\\
P & =\frac{\rho}{m_{n}}\;T,\label{mm1}\\
\frac{dP}{dr} & =-\rho\,\frac{GM}{r^{2}}\,,\label{mm2}\\
\frac{dM}{dr} & =4\pi r^{2}\rho\,. \label{m5}
\end{align}

Solving this system of equations yields the profiles of four
functions throughout the atmosphere: pressure $P(r)$, temperature
$T(r)$, baryonic density $\rho(r)$, and the enclosed mass $M(r)$.
From them any other functions of interest can be calculated. To
analyze the system it is convenient to introduce adimensional
variables normalized to the corresponding values at the core: $x=
r/R_{c}$, $m(x)= M /M_{c}$, $t(x)= T/T_{c}$, $\varrho(x)=
\rho/\rho_{c}$, and $p(x)= P/P_{c}$. Proceeding in this way,
Eqs.~(\ref{m2})-(\ref{m5}) can be rewritten
\begin{align}
\frac{dt}{dx} & =-b_{c}\frac{\varrho}{tx^{2}}\,, \label{a1}\\
p &  =t\varrho\label{a2}\,,\\
\frac{dp}{dx} &  =-c_{c}\frac{m\varrho}{x^{2}}\,, \label{a3}\\
\frac{dm}{dx} &  =d_{c}x^{2}\varrho\,, \label{a4}
\end{align}
where
\begin{equation}
b_{c}=\frac{9\bar{\varkappa}\rho_{c}L_{c}}{2\pi R_{c}T_{c}^{2}}\,,
 \ \ \ \ \ \ \
c_{c}=\frac{G m_{n}M_{c}}{R_{c}T_{c}}\,,
 \ \ \ \ \ \ \
d_{c}=\frac{4\pi\rho _{c}R_{c}^{3}}{M_{c}}\,.
\end{equation}
are adimensional constants. Thus the set of functions that solve
the previous system of equations depends only on these three
combinations of the core parameters.

By eliminating $\varrho$ in Eqs.~(\ref{a1}) and (\ref{a3}) we
arrive to
\begin{equation}
\frac{dp}{dx}=\frac{c_{c}}{b_{c}}\,mt\frac{dt}{dx}\,.\label{dp}%
\end{equation}
The last equation is easily integrated as follows:
\begin{equation}
p=\frac{t^{2}-a}{1-a}\,,\label{p}%
\end{equation}
where
\begin{equation}
a(x)=1-\frac{2b_{c}}{c_{c}\bar{m}(x)}\label{am}%
\end{equation}
and $\bar{m}(x)$ is an effective mass defined by
\begin{equation}
\int_{1}^{x}\!dx^{\prime}\,m(x^{\prime})\,t\,\frac{dt}{dx^{\prime}}\equiv\bar
{m}(x)\!\int_{1}^{x}dx^{\prime}\,t\,\frac{dt}{dx^{\prime}}\,.
\end{equation}
From Eqs.~(\ref{a2}) and (\ref{p}) we can also express the density
in terms of the temperature as
\begin{equation}
\varrho = \frac{t^{2}-a}{t(1-a)}\,, \label{rr1}
\end{equation}
which, together with Eq. (\ref{a1}), yields a first order
differential equation for $t$
\begin{equation}
\frac{dt}{dx}+\frac{b_{c}}{t^{2}x^{2}}\frac{t^{2}-a}{1-a}=0\,.
\label{dif}
\end{equation}
At $x=1$ ($R=R_{c}$) we have $\left. {dt}/{dx}\right\vert
_{x=1}=-b_{c}$, which is independent of $a(x)$. For an idealized
atmosphere, where the model would apply to the whole space, the
physical solutions would correspond to an infinite protostar where
the temperature has an asymptotic behavior for $x\gg1$, such that
the adimensional temperature tends to $t_{s}\simeq\sqrt{a}$. Thus,
the function $a(x)$ varies in the range
$1-2b_{c}/c_{c}<a<t_{s}^{2}$ for $1<x<\infty$.

Eq.~(\ref{dif}) has no analytical solution when $a$ is a function
of $x$. An approximate solution can be found by the following
procedure. Let us consider the differential equation (\ref{dif})
with $a$ constant. Then, an analytical (implicit) solution is
given by
\begin{equation}
t-1+\frac{\sqrt{a}}{2}\left[\ln\left(
\frac{t-\sqrt{a}}{t+\sqrt{a}}\right)-\ln\left(
\frac{1-\sqrt{a}}{1+\sqrt{a}}\right)\right] =\frac{b_{c}}
{1-a}\left(\frac{1}{x}-1\right)\,. \label{tem}
\end{equation}
with $1>t>\sqrt{a}$. If we replace the constant $a$ by a well
behaved function of $x$, the above expression still satisfies
Eq.~(\ref{dif}) at $x=1$. For an infinite atmosphere, a good
approximation to the exact solution is given by Eq.~(\ref{tem}),
with $a$ now a function of $t(x)$:
\begin{equation}
a(t)=1-\frac{2b_{c}}{c_{c}}-A\left(1-t\right)\,, \label{at}
\end{equation}
where $A= \left(
1-\frac{2b_{c}}{c_{c}}-t_{s}^2\right)\!/\,(1-t_{s})$. From Eq.
(\ref{at}), we see that $a(t_{s})=t_{s}^{2}$ and
$a(1)=1-2b_{c}/c_{c}$, which means that this ansatz fits the
extreme values of $a(x)$.

Once $\rho$ and $T$ are known, we can find the electron fraction
$Y_e$ as a function of $r$ by solving the implicit expression
derived in Eq.~(\ref{ye}). To examine the prediction of the model
in connection with the kick mechanism, we consider a protoneutron
star with reasonable values for its core parameters, the same as
in Ref.~\cite{BDMZ}. Thus we take
$M_{c}=M_{\odot}=1.989\times10^{30}\,\rm{kg}$,
$R_{c}=10\,\rm{km}$,
$L_{c}=2\times10^{52}\,\rm{erg}\,\rm{s}^{-1}$,
$\rho_{c}=10^{14}\,\rm{g}\,\rm{cm}^{-3}$, and $T_{c}=40\,
\rm{MeV}$. In this case,
\begin{equation}
b_{c}=1.96\,, \ \ \ \ \ \ c_{c}=3.52\,, \ \ \ \ \ \ d_{c}=0.625\,.
\end{equation}
A numerical estimation for the asymptotic value of the solution of
Eqs.~(\ref{a1})-(\ref{a4}) gives $t_{s}=0.12$. The electron
fraction takes the value $Y_e = \,0.08$ at the core.

According to Eq.~(\ref{lll}) in terms of the adimensional
functions the mean free path for the electron neutrino becomes
\begin{equation}
\lambda_{\nu_{e}}\simeq\frac{8}{\varrho\,t^{2}}\;\rm{cm}\,.
\label{fp1}
\end{equation}
Inserting the above expression in the integrand of
Eq.~(\ref{meanfp}) and evaluating the resulting integral
numerically we obtain $x_{\nu_{e}}\simeq\,2.6$ for the radius of
the $\nu_e$-sphere. This yields
$\rho\,(R_{\nu_{e}})\simeq1.3\times10^{11}\,\rm{g}\,\rm{cm}^{-3}$,
$T(R_{\nu_{e}})\simeq T_{s}=4.8\,\rm{MeV}$, and
$Y_{e}(R_{\nu_{e}})\simeq0.1$, which are in reasonable agreement
with typical values from numerical simulations\cite{BL,raffelt}.
In the same way, for the muon (tau) neutrino we get
$x_{\nu_{\mu,\tau}}\simeq\,2.2$. The profiles that correspond to
the solution of this model are given in Fig. 1.

From Eq.~(\ref{momas}) we can estimate again the intensity of the
magnetic field required for the kick in the case of
$\bar{\nu}_{\mu }\longleftrightarrow \bar{\nu}_{s}$. Besides the
temperature, density, and electron fraction profiles we also need
their logarithmic derivatives. For $T$ and $\rho$\ they are
immediately calculated from Eqs.~(\ref{a1})-(\ref{a3}), and the
corresponding results are
\begin{align}
h_{T}^{-1}&=-b_{c}\frac{\varrho}{tx}\frac{1}{r}\,,\\
h_{\rho}^{-1}+h_{T}^{-1}&=-c_{c}\frac{m}{tx}\frac{1}{r}\,.
\label{ht}
\end{align}
The formula for $h_{Y_e}^{-1}$ is obtained substituting the above
expressions into Eq.~(\ref{hy}). Now we have all the quantities
needed to determine the magnetic field from Eq.~(\ref{momas}) in
the spherical Eddington model. The result is plotted in Fig. 2.
From the curve presented in this figure we see that $B$ decreases
from a value of $8\times 10^{16}\,G $ in the core up to a minimum
of $3.8\times 10^{16}\ G$ at $r_{o_{\mu }}\simeq 16\,\rm{km}$.
From this point the magnetic field increases steadily as we
approach the surface of the neutrinosphere. In contrast with the
model used at the end of Section 5, here at $r\simeq 21\ km$ the
contributions from the geometrical and the temperature variations
compensate each other. At this radius the factor
$(1+2\bar{r}_{o_{\ell}})$ in the denominator of Eq.~(\ref{momas})
vanishes and the magnetic field takes arbitrarily large values.
In other words, no kick can be generated by this mechanism in the
regions near the surface of the $\nu_{\mu}$-sphere.

In the region between the core and the surface of the
neutrinosphere $R_{c}\leq r_{o_{\mu }}\leq R_{\nu _{\mu }}$, the
range of values allowed for the oscillation parameters is
\begin{equation}
10^{9}\rm{eV}^{2}\gtrsim \Delta m^{2}\cos 2\theta \gtrsim
10^{6}\rm{eV}^{2},
\end{equation}
For a small mixing angle this inequality translates into the
condition $30\,\rm{KeV}\gtrsim m_{s}\gtrsim 1\,\rm{KeV}$ for the
sterile neutrino mass. Small mixing is allowed by the adiabaticity
condition, which warrants an efficient flavor neutrino conversion
for $\tan ^{2}2\theta
>7\times 10^{-12}$ at $r_{o_{\mu }}=R_{c}$ and $\tan ^{2}2\theta
>3\times 10^{-9}$ at $r_{o_{\mu }}=R_{\nu _{\mu }}$. The results
for $m_s$ and the mixing angle agree with those derived in the
previous section and are compatible with a sterile neutrino that
is a viable dark matter candidate\cite{DH}.

\section{Conclusion}

In this work we have examined in detail a possible explanation of
the large drift velocities observed in pulsars in terms of an
asymmetric neutrino emission.  The asymmetry is a consequence of
the resonant neutrino conversions affected by the strong magnetic
field characteristic of protoneutron stars. We have shown that for
active-sterile neutrino oscillations this is a feasible kick
mechanism. Two conditions must be simultaneously satisfied to
generate a natal kick by this mechanism. First, the conversion has
to take place at the interior of the neutrinosphere of the active
neutrino, and second the magnetic field has to be intense enough
in order to induce a momentum asymmetry of the required magnitude.

We have made detailed calculations by means of the neutrino
distribution function in the diffusion approximation, combined
with the idea of a spherical resonance shell for neutrinos with
momentum $k$ that move in different directions relative to
$\mathbf{B}$. This scheme provides a better description of the
problem than the one formulated in terms of a single deformed
sphere acting as an effective emission surface of sterile
neutrinos. In our approach the sterile neutrinos produced in the
resonance shell move in every direction relative to $\textbf{B}$.
However, those going toward the interior of the protoneutron star
cross again the resonance region and are reconverted into active
neutrinos. These are within their own neutrinosphere and become
thermalized. As a consequence, only the outgoing sterile
neutrinos contribute to the total momentum emitted by the
protostar.

Two opposite effects have to be taken into account when computing
the fractional momentum asymmetry. One is purely geometric, and
comes from the difference in the areas of the semispherical
emission surface for neutrinos moving in opposite directions with
the same momentum magnitude. The other is due to the radial
temperature gradient, by which sterile neutrinos produced at
different depths have unequal energy. In general both
contributions do not compensate and thus we obtain a non null
fractional asymmetry in the total momentum. Explicit results have
been obtained for two simple models of a protostar, with the
matter background assumed to be composed by nonrelativistic
nucleons and degenerated electrons. One of them, which we worked
with in detail, is the Eddington model for a spherical neutrino
atmosphere.

Magnetic fields of the order of $10^{16}\!-\!10^{17}\,\rm{G}$ are
needed to reproduce the observed pulsar velocities. At first sight
these fields could seem rather large, compared with the estimated
values at the surface of a protoneutron star
($B_s\sim10^{13}\,\rm{G}$), but in fact intensities as high as
$B_c\sim10^{18}\,\rm{G}$ are possible at the core. A given
parametrization for the magnetic field at the interior of a
protoneutron star is\cite{manka}
\begin{equation}
B=B_{s}+B_{c}\left[1-e^{-\beta\left(\rho/\rho_{s}\right)^{\gamma}
}\right],
\end{equation}
with $\beta\simeq10^{-5}$ and $\gamma\simeq3$. By adopting this
profile as an upper boundary for the magnetic field in the
protostar, we can see that the required field remains well below
this boundary in most of the region within the neutrinosphere. In
addition, the oscillation parameters are compatible with the
allowed region for sterile neutrinos to be warm dark matter,
leaving the mechanism as an attractive possibility to explain the
 proper motion of pulsars.

\section{Acknowledgments}

This work was partially supported by CONICET-Argentina, CONACYT-
M\'{e}xico, and DGAPA-UNAM (M\'{e}xico) under grant
PAPIIT-IN109001.

\pagebreak

\section*{Figure captions}

FIG. 1: The different profiles normalized to the corresponding
value at the core for the Eddington model discussed in Section VI.
\vskip 0.5cm \noindent
 FIG. 2: The magnetic field $B$ required to
produce the kick by the $\bar\nu_\mu\rightarrow \bar\nu_s$
resonant conversion for the Eddington model discussed in Section
VI, normalized to the field required at the core, $ B_{c}$.

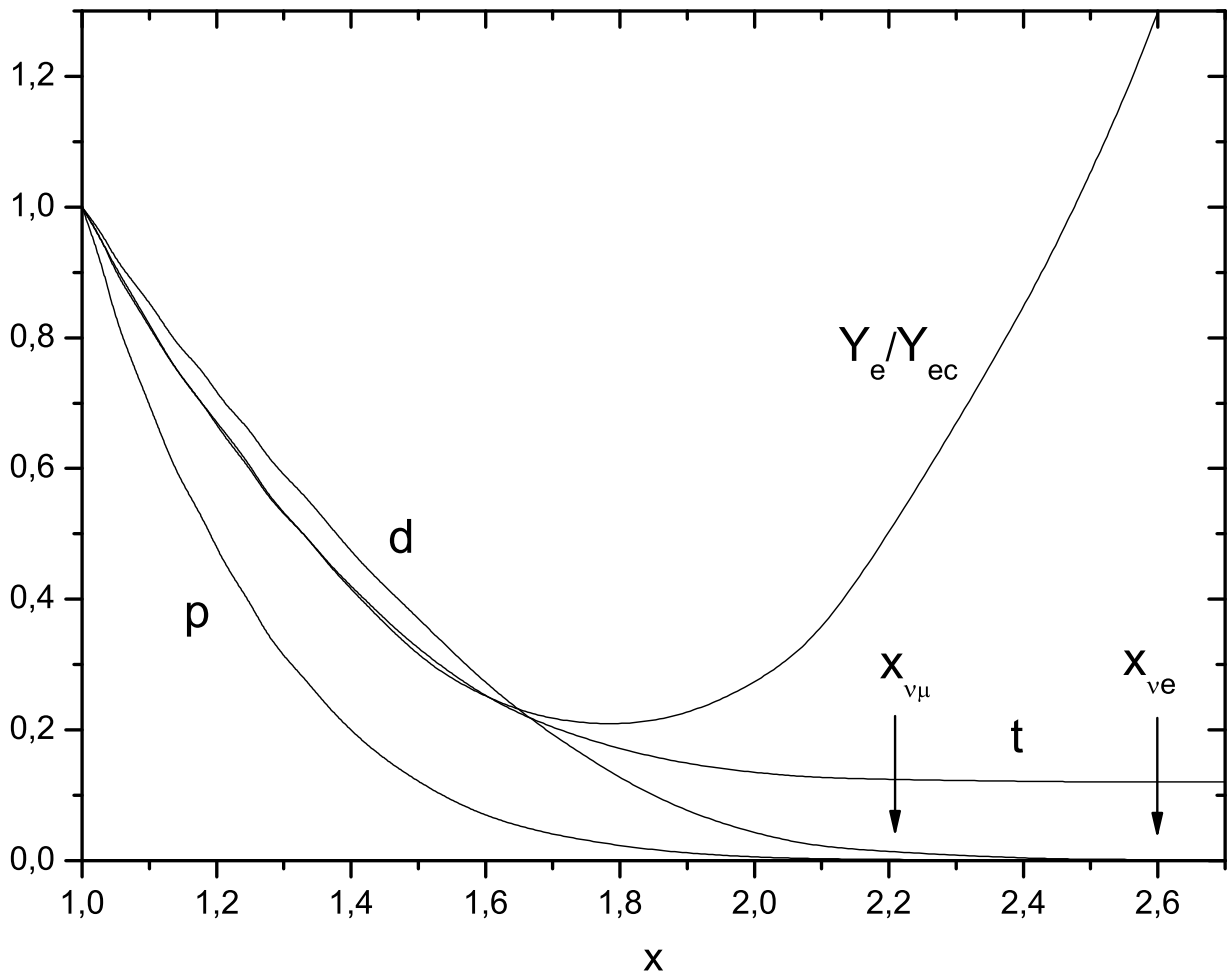
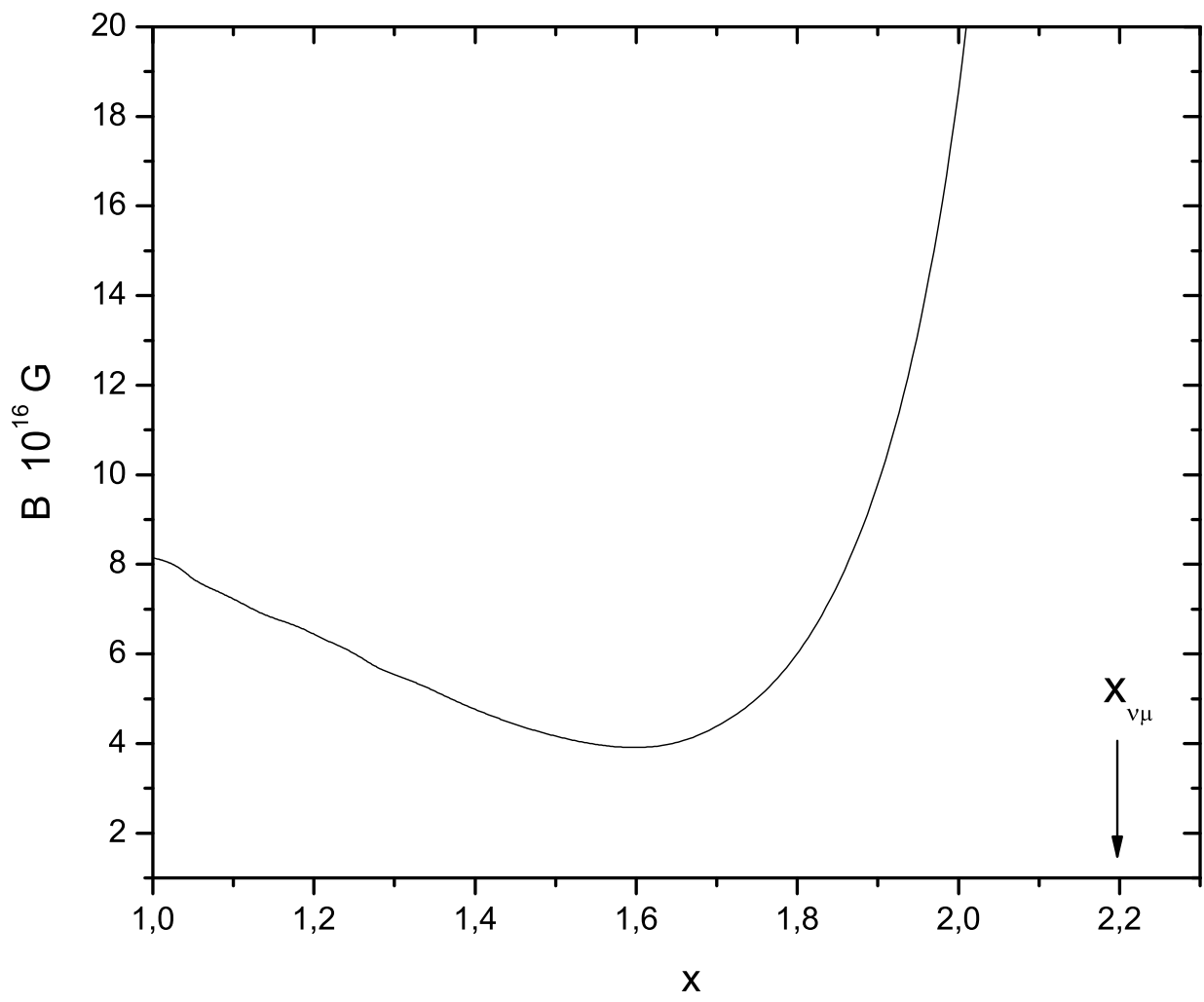
\end{document}